\documentstyle[epsfig, twocolumn]{esapub}

\def\lsim{\raise0.3ex\hbox{$<$}\kern-0.75em{\lower0.65ex\hbox{$\sim$}}}
\def\gsim{\raise0.3ex\hbox{$>$}\kern-0.75em{\lower0.65ex\hbox{$\sim$}}}
\begin{document}

\setlength{\parindent}{0pt}
\setlength{\parskip}{ 10pt plus 1pt minus 1pt}
\setlength{\hoffset}{-1.5truecm}
\setlength{\textwidth}{ 17.1truecm }
\setlength{\columnsep}{1truecm }
\setlength{\columnseprule}{0pt}
\setlength{\headheight}{12pt}
\setlength{\headsep}{20pt}
\pagestyle{esapubheadings}
\title{\bf HIGH $z$ SUPERNOVAE WITH THE NGST}

\author{{\bf T.~Dahl\'{e}n, C.~Fransson} \vspace{2mm} \\
Stockholm Observatory, S-133 36 Saltsj\"{o}baden, Sweden \\
Phone +46 8 16 44 67/69, fax +46 8 717 47 19 \\
e-mail tomas@astro.su.se, claes@astro.su.se}
\maketitle
\begin{abstract}
We use different assumptions for the combination of dust extinction and star formation up to redshifts $z$ \gsim ~ 5, as well as detailed modeling of supernova properties, to estimate the number of supernovae that should be observable with various instruments, in particular the NGST. In the model we use realistic light curves and spectral shapes that evolve with time for the different types supernovae.

We find that the NGST should be able to detect several tens of core collapse SNe in a single detection if the observational limit in the range 1-5 $\mu$m is $\sim$ 1 nJy and the field is 16 arcmin$^2$. We also estimate the observable number of Type Ia supernovae. Due to the time delay between the formation of the progenitor star and the explosion of the supernova, there are additional aspects that have to be considered when interpreting observational rates of these supernovae. \vspace {5pt} \\
  Key~words: high redshift supernovae.
\end{abstract}
\section{INTRODUCTION}
The evolution of the star formation rate is reflected in the cosmic supernova rate (SNR). It should therefore, in principle, be possible to use SN observations to distinguish between star formation scenarios. Even more important, core collapse SNe, i.e. Types II and Ib/c, provide a direct probe of the metallicity production with cosmic epoch. Due to their importance as probes for early cosmological epochs, studies of high $z$ SNe is one of the main goals for the Next Generation Space Telescope (NGST) (Stockman \& Mather \cite{stockman}).

In this paper we present estimates for the expected number of observable supernovae and discuss some of the complications entering the analysis. Previous studies include e.g. Madau et al. (\cite{madau98}), Madau (\cite{madau97}), Ruiz-Lapuente \& Canal (\cite{ruiz}), Yungelson \& Livio (\cite{yungelson}). With respect to these our work differs in that we include information about the light curve, which allows us to predict the simultaneously observable number of SNe. We divide the SNe into different types with maximum absolute magnitudes accordingly. We also calculate the counts for different filters and include information about the estimated redshifts of the detected SNe.

Section 2 describes our model. Results are presented in section 3. In section 4 we discuss the influence of dust. Further discussion follows in section 5. Conclusions are given in section 6. We assume a cosmology with $H_0=50~{\rm km~s^{-1}~Mpc}^{-3}$ and $q_0$=0.5.
\section{THE MODEL}
\label{sec:model}
\begin{figure}[t]
\begin{center}
\leavevmode
\centerline{\epsfig{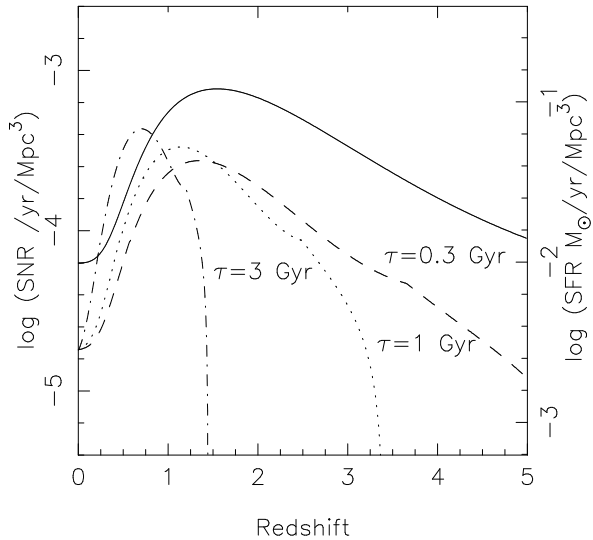}}
\end{center}
\caption{\em Supernova rates for the model with dust extinction $E_{B-V}$=0.1. The solid line represent the rate of core collapse supernovae. Dashed, dotted and dash-dotted lines represent the rate of Type Ia supernovae with time delays $\tau$=0.3, 1.0 and 3.0 Gyr respectively. The rates of Type Ia SNe have been normalized so that the ratio between the core collapse and the Type Ia rates matches the locally observed value 3.5. Solid line together with scales on the right y-axis show the star formation rates.}
\label{fig1}
\end{figure}

A problem when using luminosity densities from high $z$ galaxies, to calculate the SFR, is that these can be matched to wide range of SFRs, spanning from strongly peaked to flat or even increasing at $z\gsim$1, by adjusting the assumed extinction due to dust. Madau (\cite{madau97}) finds, by using luminosity densities in three bands (UV, optical and NIR) that a universal extinction $E_{B-V}$=0.1 with SMC-type dust makes the best fit between a proposed SFR and the observations. This SFR shows a pronounced peak at 1$<z<$2, see Figure \ref{fig1}. This shape is predicted in a scenario where galaxies form hierarchically (e.g. Cole at al. \cite{cole}).

In this paper we calculate the supernova rates by using the star formation rate and extinction $E_{B-V}$=0.1, derived by Madau (\cite{madau97}). We also comment on how a higher extinction affects the rates, and if it is possible to use supernova counts to make predictions about the amount of dust.

Core collapse supernovae, which are thought to originate from short lived massive stars, have an evolution that closely follows the shape of the SFR. Assuming an immediate conversion of these stars to supernovae renders a multiplicative factor between the star formation and the supernova rates.
\begin{equation}
{\rm SNR=}k\times{\rm SFR},
\end{equation}
where $k$ depends on the IMF and the mass range of the progenitors.

The uncertainties concerning the life-time of Type Ia progenitors make the relation between the rate of these supernovae and the star formation more difficult to model. We treat the life-time $\tau$ of the progenitor WD as a free parameter which is allowed in the range 0.3$<\tau<$3 Gyr.

In order to derive the number of SNe to different limiting magnitudes, the apparent magnitude of the SNe are calculated from
\begin{equation}
m_f(z,t,i)= M_f(t) + \mu(z) + K_f(z,t) + A_g + <A_{i,f}>
\end{equation}
Here $M_f(t)$ is the absolute magnitude in a filter $f$ at time $t$ relative to the peak of the light curve, $\mu (z)$ is the distance modulus, $K_f(z,t)$ gives the K-correction, $A_g$ is the Galactic absorption, and $<A_{i,f}>$ is the absorption in a parent galaxy with inclination $i$. The core collapse SNe are divided into five different types, i.e. Ib/c, IIP, IIL, SN1987A-type and IIn, with absolute magnitudes accordingly. We model the spectra of the different SN types by blackbody curves with cutoffs at short wavelengths. The spectra are also set the evolve with time.

An important feature of our model is that the SNe are distributed over the light curve, and are given absolute magnitudes accordingly. With this procedure we obtain the simultaneously observable number of supernovae. In order to actually detect the SNe, a second observation has to be made after an appropriate time has passed.
The effect of the time delay between the formation of the progenitor star and explosion of Type Ia SNe shifts the peak of the rate of these SNe towards lower $z$ than compared to the core collapse SNe. In Figure \ref{fig1} we show the rates for core collapse and Type Ia SNe. Three different values for the life-time of Type Ia progenitors are shown, $\tau$=0.3, 1 and 3 Gyr. The rates of Type Ia's are normalized so that the fraction between core collapse and Type Ia rates matches the observed value at z=0.
\section{RESULTS}
\label{sec:results}
\begin{figure*}
\begin{center}
\leavevmode
\centerline{\epsfig{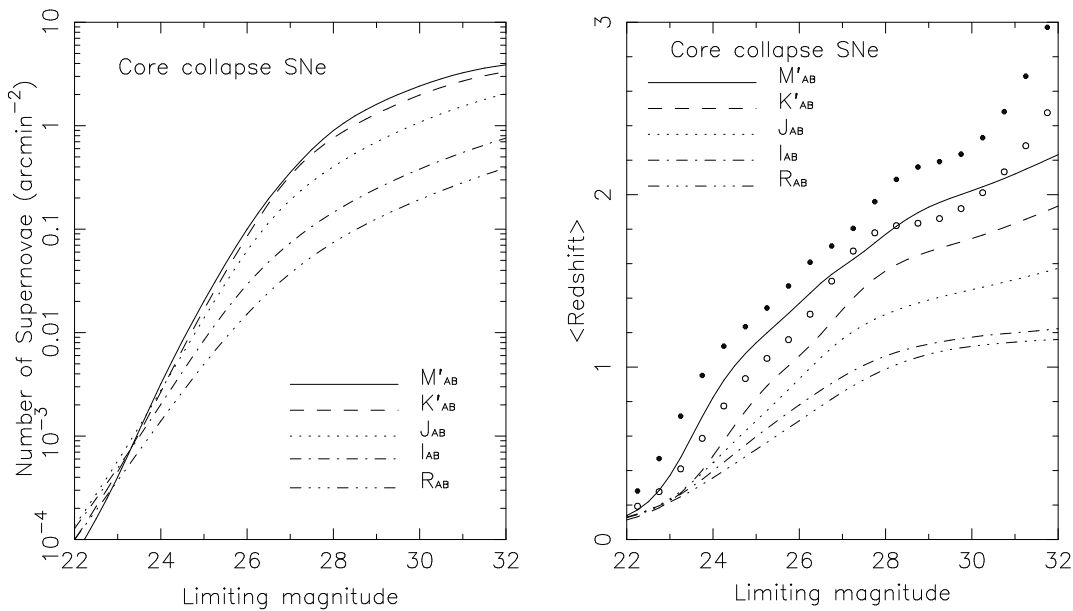}}
\end{center}
\caption{\em Left panel: the number of core collapse supernovae per arcmin$^2$ that can be detected down to different limiting magnitudes. Results for M$^{\prime}$, K$^{\prime}$, J, I and R bands are shown as full, dashed, dotted, dash-dotted and dash-triple-dotted lines, respectively. Right panel: the mean redshift of the SNe (same representation as left panel). Filled and open dots represents the redshifts in magnitude bins of $\Delta$m=0.5 for the M$^{\prime}$ and K$^{\prime}$ bands. Note that AB-magnitudes are used.}
\label{fig2}
\end{figure*}
\subsection{Core Collapse Supernovae}
Figure~\ref{fig2} shows the predicted number of simultaneously observable core collapse SNe per arcmin$^2$ in the M$^{\prime}$ (centered on 4.2$\mu$m with $\Delta \lambda=\pm 0.8 \mu$m), K$^{\prime}$ (2.1$\mu$m), J (1.2$\mu$m), I and R filters for different limiting magnitudes (given in the AB system). Also shown are the mean redshifts of the SNe at the different limits and the mean redshifts in $\Delta$m=0.5 magnitude bins for the K$^{\prime}$ and M$^{\prime}$ filters.
   
The number of observable high $z$ SNe increases considerably more at faint limiting magnitudes for the filters that probe longer wave-lengths. This K-correction effect is enhanced by fact that a majority of the core collapse SNe has a spectral cutoff at short wave-lengths, causing these SNe to drop out of the R, I, J, K$^{\prime}$ and M$^{\prime}$ filters at $z \sim$0.9, $\sim$1.4, $\sim$2.7, $\sim$5.5 and $\sim$10, respectively.

The NGST is planned to have a detection limit of $\sim$1 nJy in the J and K$^{\prime}$ bands and $\sim$3 nJy in the M$^{\prime}$ band for a 10$^4$s exposure with S/N=10 and R=3 (NGST Simulation Group, 1998). These fluxes correspond to AB-magnitudes of 31.4 and 30.2, respectively. With these limits we find that the K$^{\prime}$ band yields the highest predicted number of observable SNe. In a 4$\times$4 arcmin field we predict $\sim$ 47 simultaneously detectable core collapse SNe in the K$^{\prime}$ band. The numbers for the J and M$^{\prime}$ bands are $\sim$ 28 and $\sim$ 41, respectively. The mean redshift of the SNe are $<z>\sim$ 1.9 in the K$^{\prime}$ band and $<z>\sim$ 2.1 in the M$^{\prime}$ band. The mean redshift in the faintest observable magnitude bin ($\Delta$m=0.5 mag) for both these filters are $<z>\sim$ 2.3. The number of simultaneously observable SNe with $z>2$ is $\sim$ 17 for both the K$^{\prime}$ and the M$^{\prime}$ filter. At even higher $z$, the M$^{\prime}$ filter is preferred due to the effects of K-correction. For SNe with redshifts $z>5$ and $z>9$ we predict 1.1 and 0.09 SNe in the M$^{\prime}$ band, while the numbers for the K$^{\prime}$ band are 0.3 and 0.02 (using field and detection limits as above). These high $z$ estimates are based on a supernova rate derived by extrapolating the rate from $z\lsim$ 5. This makes the predicted numbers somewhat uncertain. However, the M$^{\prime}$ band should detect a factor $\gsim$ 3-5 times more SNe than the K$^{\prime}$ band at these $z$, independent of the actual rate. We stress that it is necessary to use filters that sample wave-lengths long ward of the (redshifted) spectral cutoff of the SNe spectra in order to detect SNe at high $z$. Note that the number of observable SNe with $z$ above $\sim$ 2-4 may be increased by the effects of gravitational magnification (Marri \& Ferrara \cite{marri}).

An HDF-like observation, covering 5 arcmin$^2$ with observational limits of I$\lsim$29 and R$\lsim$29, should contain $\sim$ 1.2 and $\sim$ 0.6 core collapse SN in each field. The ESO Very Large Telescope (VLT) reaches R$\sim$26.8 using FORS in 1800s exposure (Moorwood \cite{moorwood}). With a field of 6.8'$\times$6.8', one should detect $\sim$ 1.5 core collapse SNe. The mean redshift of these SNe is predicted to be $<z>\sim$ 0.8. About 30\% and 0.8\% of these have $z>$1 and $z>$2, respectively. Reaching a magnitude deeper would raise the detected number to $\sim$ 3 and increase the mean redshift to $<z>\sim$ 1. With this limit 40 \% and 2 \% of the SNe have redshifts $z>$1 and $z>$2. Using the VLT infrared ISAAC instrument, which is specified to reach K$^{\prime}>$24.8 in a 2.5'$\times$2.5' field (3600s exposure) (Moorwood \cite{moorwood}), will yield around 0.07 detected SNe per observation ($<z>\sim$ 0.8). Reaching a magnitude deeper increases the predicted number of observable SNe to 0.4 per field ($<z>\sim$ 1). In the last example about 47\% and 4\% of the SNe have $z>$1 and $z>$2, respectively. These estimates show that ground based instruments can not compete with the NGST when it comes to detecting high $z$ supernovae.
 
So far we have discussed the number of SNe that are simultaneously observable during one search (requiring two observations to actually detect the SNe). Prolonging the time coverage of a specific field increases the observed number of SNe. With 4-6 observations per year, yielding a full coverage of the light curves, we estimate a total of $\sim$ 70 SNe to be detected per field with the NGST in the K$^{\prime}$ filter.

Compared to earlier published estimates we predict greater numbers for the counts. With the same extinction, Madau et al. (\cite{madau98}) predict $\sim$7 SNe per NGST field per year in the range 2$<z<$4. Our calculations result in $\sim$23 (both K$^{\prime}$ and M$^{\prime}$). The main reason is that we include SNe observed over the whole light curve.
\subsection{Type Ia Supernovae}
\begin{figure*}
\begin{center}
\leavevmode
\centerline{\epsfig{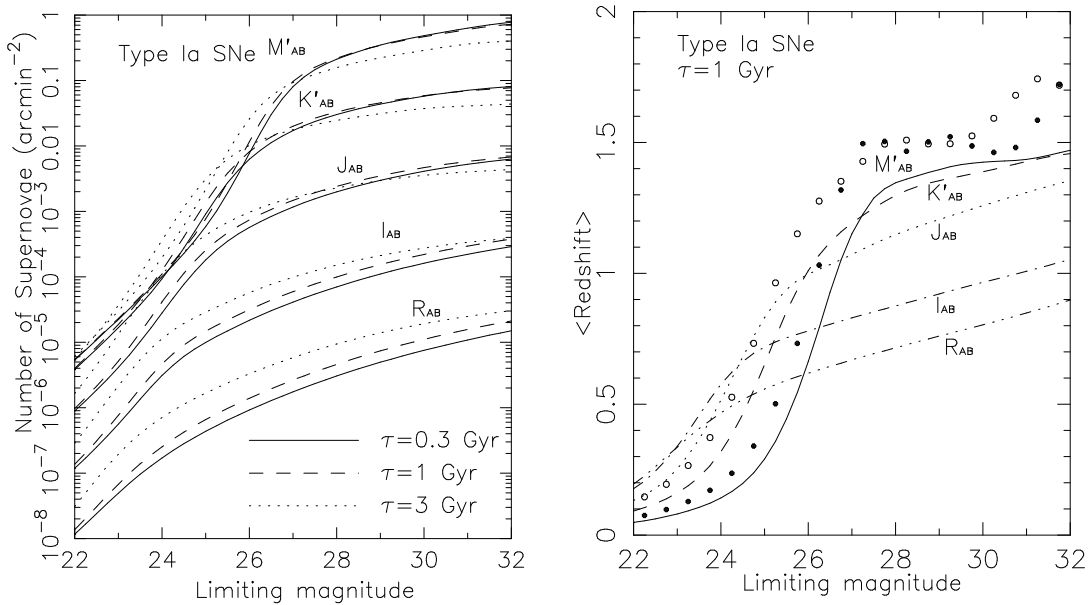}}
\end{center}
\caption{\em Left panel: the number of predicted Type Ia SNe per arcmin$^2$ in the M$^{\prime}$, K$^{\prime}$, J, I and R filters for different limiting magnitudes. The full, dashed and dotted lines represents $\tau$=0.3, 1.0 and 3.0 Gyr. The left hand scaling is valid for the M$^{\prime}$ counts. The counts in other filters are off set by factors 10$^{-1}$, 10$^{-2}$, 10$^{-3}$and 10$^{-4}$, respectively. Left panel: the lines show the mean redshifts of the SNe at the different limits for the $\tau$=1 Gyr model. Filled and open dots show the mean redshift in magnitude bins ($\Delta$m=0.5) for the M$^{\prime}$ and K$^{\prime}$ bands, respectively. Note that AB-magnitudes are used.}
\label{fig3}
\end{figure*}
Figure \ref{fig3} shows the detectable number of Type Ia SNe down to different limiting magnitudes for $\tau$=0.3, 1 and 3 Gyr in the different filters. Also shown is the mean redshift for the model with $\tau$=1 Gyr.

These results show that a change in the delay time, $\tau$, introduces a fairly large dispersion in the predicted number of observable SNe. For example, with the NGST K$^{\prime}$ detection limit $\sim$ 12 SNe are predicted for the two low $\tau$ values and $\sim$ 7 for $\tau$=3 Gyr. This may seem as a possible observational probe to distinguish between different progenitor models. However, in next section we show that this use is hampered by the uncertainties introduced when considering alternative extinction models.
\section{DEPENDENCE ON DUST}
\label{sec:dust}
There are several claims that the extinction $E_{B-V}$=0.1, as argued for by Madau~(\cite{madau97}), and used here, is too low. Observations of high $z$ galaxies, as well as the far-infrared extra-galactic background (e.g. Meurer~\cite{meurer97}; Rowan-Robinson et al.~\cite{rowan}; Sawicki \& Yee~\cite{sawicki}; Ellingson et al.~\cite{ellingson}; Soifer et al.~\cite{soifer}; Burigana et al.~\cite{burigana}), indicate that an extinction of $E_{B-V}\sim$0.3 could be more likely. With a high extinction it is possible that the observed peak in the UV-luminosities from the high $z$ galaxies is illusive and that the star formation history is more compatible with a monolithic collapse scenario. To mimic the case of a monolithic collapse, Madau (\cite{madau97}) presents a test case where the extinction is adjusted so that a proposed monolithic SFR matches the observations. In this model he finds an extinction law that increases with redshift as $E_{B-V}$=0.011(1+$z$)$^{2.2}$.

We have run our model with the monolithic extinction law in order to examine how a higher absorption a high redshifts affects the predicted rates.
\subsection{Core collapse SNe}
The rate of exploding high $z$ core collapse SNe increases in proportion to the assumed increase in star formation in the high extinction model. Due to the effects of the dust, the observable number of SNe will, however, be similar to the numbers found in the low extinction model. This is hardly surprising, because the observed supernova rate is, except for the K-correction, proportional to the observed star formation rate, which (by assumption) is the same in the two models. With the NGST limits, the number of detected SNe in a field increases from $\sim$47 in the low dust scenario to $\sim$54 in the high dust scenario, using the K$^{\prime}$ band.

Since the K-corrections increase with redshift, it may, however, be possible to use a high redshift subsample of the observed SNe in the K$^{\prime}$ or M$^{\prime}$ band as a probe for the amount of dust. Using NGST limits, and only counting SNe with $z>$ 2 in the low and high dust models, results in $\sim$ 17 and $\sim$ 32 detections, respectively (in both bands).
\subsection{Type Ia SNe}
The increased star formation at high redshifts in the monolithic model makes the rate of high $z$ Type Ia SNe sensitive to the choice of formation redshift, $z_F$. Allowing the formation redshift to be in to range 5 $<z_F<$ 9, together with the uncertainty in progenitor lifetime  $\tau$, leads to a wide dispersion in the estimated counts of Type Ia SNe. Using the NGST limits results in $\sim$5-36 simultaneously observable Type Ia SNe in the K$^{\prime}$ filter. A better understanding of the properties of the dust seems necessary if the rates are to be used as probes for different progenitor scenarios. 
\section{DISCUSSION}
The actual amount of extinction has a large influence on the cosmic star formation rate above redshifts of the peak at $z\sim$1.5. This affects the intrinsic number of supernovae and therefore also the production and enrichment of metals by core collapse SNe throughout history. The number of core collapse supernovae as observed from Earth is, however, less dependent on the amount of extinction, due to the fact that high intrinsic star formation rates are combined with high absorption as to match the same observational data. A more pronounced change in the derived rates would emerge if the
observed UV luminosities from the high $z$ galaxies under-predicts the star formation due to other effects than extinction. This could be the case if a large part of the star formation takes place in low surface brightness galaxies, or if selection effects biases against the detection of galaxies with increasing redshift (e.g. Hu et al. \cite{hu}; Ferguson \cite{ferguson}).

The rates of simultaneously detectable SNe results from events over the whole light curve. Using Type Ia's as standard candles requires observations at the peak of the light curve, i.e that a first detection is made at the rising part of the light curve.
Our calculations predict that the NGST will be able to simultaneously detect $\sim$ 0.6 Type Ia's on the rising part of the light curve, of which $\sim$ 0.4 and $\sim$ 0.1 has redshifts $z>$1 and $z>2$, respectively. We have used the hierarchical model with extinction $E_{B-V}$=0.1, and progenitor life-time $\tau$=1 Gyr, in these examples. The dependence of these numbers on the progenitor life-time is, however, strong.

The major sources of uncertainty in our calculations is most likely to be connected with the treatment of the dust extinction and the assumed mass range of the SN progenitors. In Dahl\'{e}n \& Fransson (\cite{dale}) we discuss these, and other possible sources of uncertainty. In this paper we also go deeper into the details of our model and include discussions on detectability, metal production and the influence of cosmological models on the derived rates.
\section{CONCLUSIONS}
We have found that the predicted number of core collapse SNe is rather insensitive to the assumed star formation scenario, as long as the star formation is calculated to match the same observed luminosity densities. Different star formation models may, however, be tested by using high redshift subsamples of core collapse SNe. 

When it comes to the observable rates of Type Ia SNe, we find that these are more sensitive to the star formation and dust modeling. This is due to the fact that the Type Ia's are less linked to the environment where their progenitors were formed. The uncertainty in the life-time of the progenitors, combined with the sensitivity on the rates to the onset of star formation in monolithic-like models, contributes to the difficulty with using Type Ia's as probes for either different star formation scenarios, or progenitor models.

The number of core collapse SNe that could simultaneously be detected with the NGST down to 1 nJy (m$_{AB}=$31.4) in a 4$\times$4 arcsec field is calculated to be $\sim$ 47 in the K$^{\prime}$ band. In order to detect SNe with $z\gsim$ 2, long wave-length filters (i.e. M$^{\prime}$) should be used. This is due to K-correction. The effect of the K-correction is enhanced by the spectral cutoff, causing a majority of the SNe to drop out of the shorter wave-length filters at relatively lower redshifts. The high number predicted for the counts implies that searches for these SNe will be very important when it comes to the understanding of the star formation and the nucleosynthesis in the early Universe. 

The number of simultaneously detectable Type Ia SNe with the NGST limits are $\sim$5-35, depending on model. Additional uncertainties widen this range even more. Of the simultaneously observable Type Ia's, only about 5\% are at the rise of the light curve.
{}
\end{document}